\documentclass[reprint, superscriptaddress, secnumarabic, amssymb, nobibnotes, aps, prl]{revtex4-1}

\usepackage{graphicx}
\usepackage{epstopdf}
\usepackage[T1]{fontenc}
\usepackage[latin9]{inputenc}
\usepackage{amsbsy}
\usepackage{gensymb}
\begin{document}

\title{\textrm{Investigations of the superconducting ground state of Zr$ _{3} $Ir: Introducing a new noncentrosymmetric superconductor}}
\author{Sajilesh.~K.~P.}
\affiliation{Department of Physics, Indian Institute of Science Education and Research Bhopal, Bhopal, 462066, India}
\author{D.~Singh}
\affiliation{Department of Physics, Indian Institute of Science Education and Research Bhopal, Bhopal, 462066, India}
\author{P.~K.~Biswas}
\affiliation{ISIS Facility, STFC Rutherford Appleton Laboratory, Harwell Science and Innovation Campus, Oxfordshire, OX11 0QX, UK}
\author{Gavin.~B.~G.~Stenning}
\affiliation{ISIS Facility, STFC Rutherford Appleton Laboratory, Harwell Science and Innovation Campus, Oxfordshire, OX11 0QX, UK}
\author{A.~D.~Hillier}
\affiliation{ISIS Facility, STFC Rutherford Appleton Laboratory, Harwell Science and Innovation Campus, Oxfordshire, OX11 0QX, UK}
\author{R.~P.~Singh}
\email[]{rpsingh@iiserb.ac.in}
\affiliation{Department of Physics, Indian Institute of Science Education and Research Bhopal, Bhopal, 462066, India}

\date{\today}
\begin{abstract}
\begin{flushleft}
\end{flushleft}

Superconductivity in materials whose crystal structure lacks inversion symmetry is a prime candidate for unconventional superconductivity. A new noncentrosymmetric compound Zr$_{3}$Ir crystallizes in a tetragonal $ \alpha $-V$ _{3} $S structure. The magnetization, specific heat and muon spin rotation confirm s-wave superconductivity, having a transition temperature T$_{c}$ = 2.3 K. Muon spin relaxation confirms the preservation of time reversal symmetry in the superconducting ground state.
\end{abstract}

\maketitle
\section{Introduction}
Noncentrosymmetric superconductors have attracted considerable attention recently in both theoretical and experimental condensed matter physics.The lack of inversion symmetry in these materials allow an antisymmetric spin-orbit coupling (ASOC) which can lift the degeneracy of the conduction band electrons and cause the superconducting Cooper pairs to contain an admixture of spin-singlet and spin-triplet states\cite{EBA, rashba1,rashba2,rashba3,vm,kv,fujimoto}. This mixed pairing may leads to superconductors with exotic properties, which are generally not observed in conventional superconductors, e.g. high upper critical field, time reversal symmetry breaking (TRSB), topologically protected edge states and anisotropic superconducting gap \cite{edg1,edg2,Bauer2004}. Time reversal symmetry breaking is rarely observed phenomenon phenomenon and only has been observed in a very few unconventional superconductors \cite{Sr2RuO4,OpticalSr2RuO4,UPt3,UThBe,PrOsSb,PrPtGe1,LaNiGa2,LuRhSn,Re}. Noncentrosymmetric materials are prime members to host TRS breaking due to admixed ground state and its mixing ratio tunability using the strength of ASOC. Several noncentrosymmetric materials have been investigated to search unconventional superconductivity, among which LaNiC$_{2} $ \cite{LNC1}, La$ _{7} $Ir$ _{3} $\cite{LI}, Re$_{6} $X (X = Zr, Hf, Ti) \cite{RZ3,RT,RH2}, Re$ _{24} $Ti$ _{5} $ \cite{RT2}, locally noncentrosymmetric SrPtAs \cite{SPA} are reported to show the presence of spontaneous static or quasistatic magnetic fields below the superconducting transition indicating a broken time reversal symmetry in the superconducting state. At the same time, many NCS superconductors are reported to show a conventional superconducting ground state \cite{LPS,LRS,RTa,LMP,CIS,LS2,LS3,RhMoN,SrAuSi}. Hence, it is important to understand the role ASOC and electron correlations on the parity mixing in these materials to clearly understand the presence/absence of TRSB. The only a small number of noncentrosymmetric superconductors which exhibit TRSB makes it difficult to determine the roles of ASOC. Therefore it is crucial to discover and characterize new superconductors whose crystal structure lack inversion symmetry. 

In this paper, we study superconducting properties of Zr$_{3}$Ir containing heavy (4d and 5d) elements which crystallize into noncentrosymmetric tetragonal $ \alpha $-V$ _{3} $S structure \cite{zi}. To the best of our knowledge, no other noncentrosymmetric superconductors with  $ \alpha $-V$ _{3} $S structure has been studied in detail. The noncentrosymmetric crystal structural along with the presence of heavy transition metals makes Zr$_{3}$Ir an interesting candidate to investigate the superconducting ground state. Detailed magnetization, heat capacity, and muon measurements suggest s-wave superconducting ground state with preserved time reversal symmetry.

\section{Experimental Details}
Polycrystalline sample of Zr$ _{3} $Ir was prepared by conventional arc melting technique in which the constituent elements (Re powder 99.99\% \textit{Alfa Aesar}; Zr slug 99.99\% \textit{Alfa Aesar} ) were taken in the stoichiometric ratio and melted in a water-cooled copper hearth under high purity argon gas. The elements were melted to form small ash coloured button of Zr$ _{3} $Ir, which was flipped and remelted several times for homogeneity of the sample.

Characterization of the crystal structure and phase purity of the sample was done by room temperature X-ray diffraction measurements using PANalytical diffractometer equipped with CuK$ _{\alpha} $ radiation ($ \lambda $ = 1.5406 \text{\AA}). Magnetization and AC susceptibility measurements were done using Quantum Design superconducting quantum interference device (MPMS 3, Quantum Design). Specific heat and resistivity measurements of the sample was done in zero field as well as applied field (resistivity) using Quantum Design physical property measurement system (PPMS, Quantum Design, Inc.). The $ \mu $SR experiments were carried out using 100\% spin-polarized pulse muon beam at the ISIS facility, STFC Rutherford Appleton Laboratory, Didcot, United Kingdom. Both longitudinal and transverse field measurements were carried out with detectors which can be aligned accordingly while sample was mounted on a high purity silver holder. Correction coils were used to neutralize the stray fields at the sample position.

\section{Results and Discussions}
 \begin{figure} 
	\includegraphics[width=9. cm, height=6. cm, origin=b]{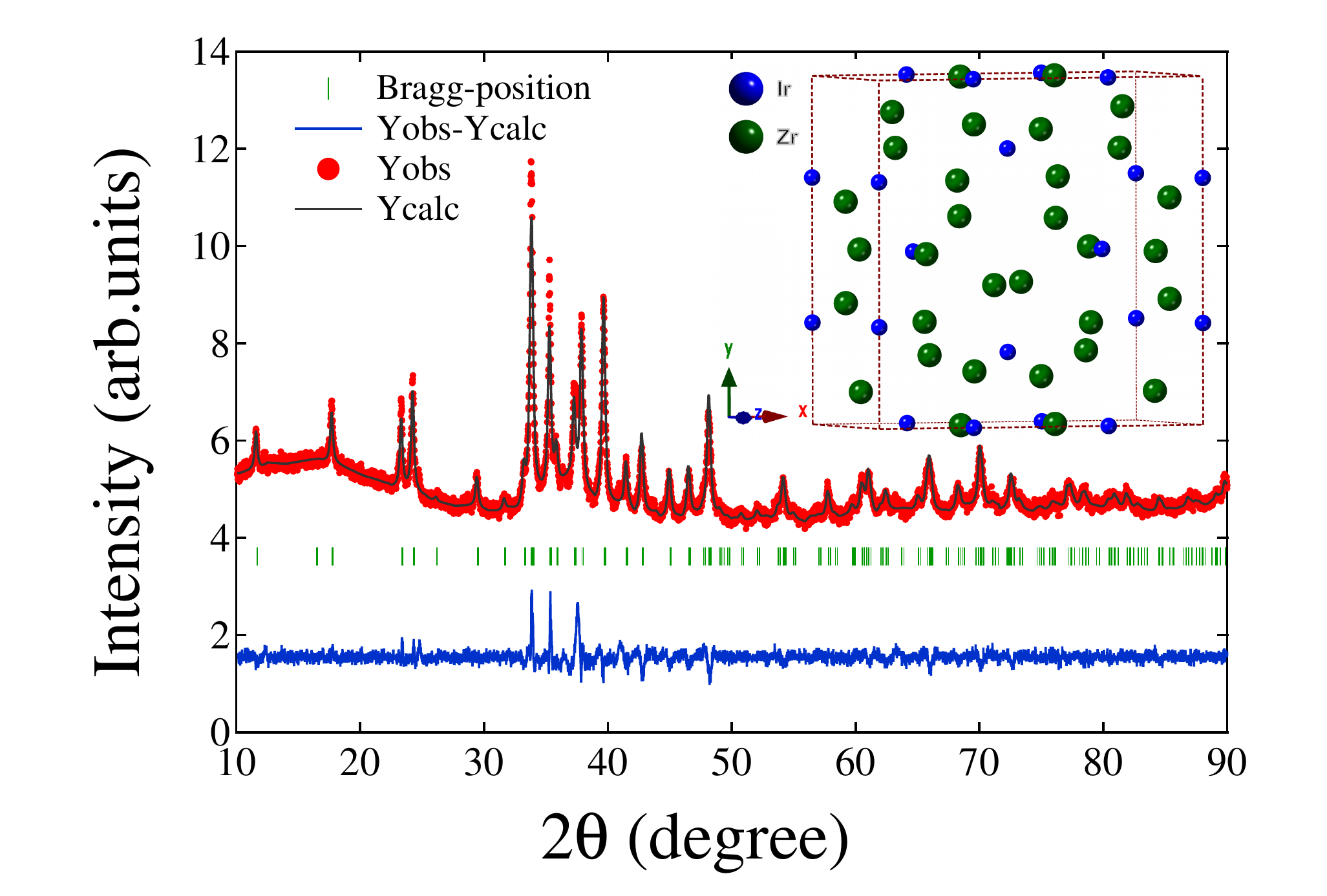}
	\caption{Powder XRD pattern of the sample recorded at ambient temperature using Cu$K_{\alpha}$ radiation (red line). The Rietveld refined data for the noncentrosymmetric space group I-42m (121) is shown as a dotted black line. The green vertical lines shows the calculated reflection positions. The inset displays a unit cell of Zr$ _{3} $Ir. }
	\label{fig1}
\end{figure}
\subsection{Crystallography}
Figure \ref{fig1} shows X-ray diffraction pattern of Zr$ _{3} $Ir. The structural refinement of the data was carried out using Fullprof software. It confirms that sample crystallized in the noncentrosymmetric tetragonal $ \alpha $-V$ _{3} $S structure (space group no - 121) with no impurity. The lattice parameters obtained from refinement are a = b = 10.788(4) \text{\AA}, c = 5.602(2) \text{\AA}, consistent with earlier reports \cite{zi}. Noncentrosymmetric nature of crystal structure can be seen from the arrangement of Zr atoms (Fig. \ref{fig1} inset). A detailed description regarding the crystal structure can be seen in Ref. \cite{zi}.

\subsection{Resistivity}
\begin{figure} 
\includegraphics[width=9. cm, height=6. cm, origin=b]{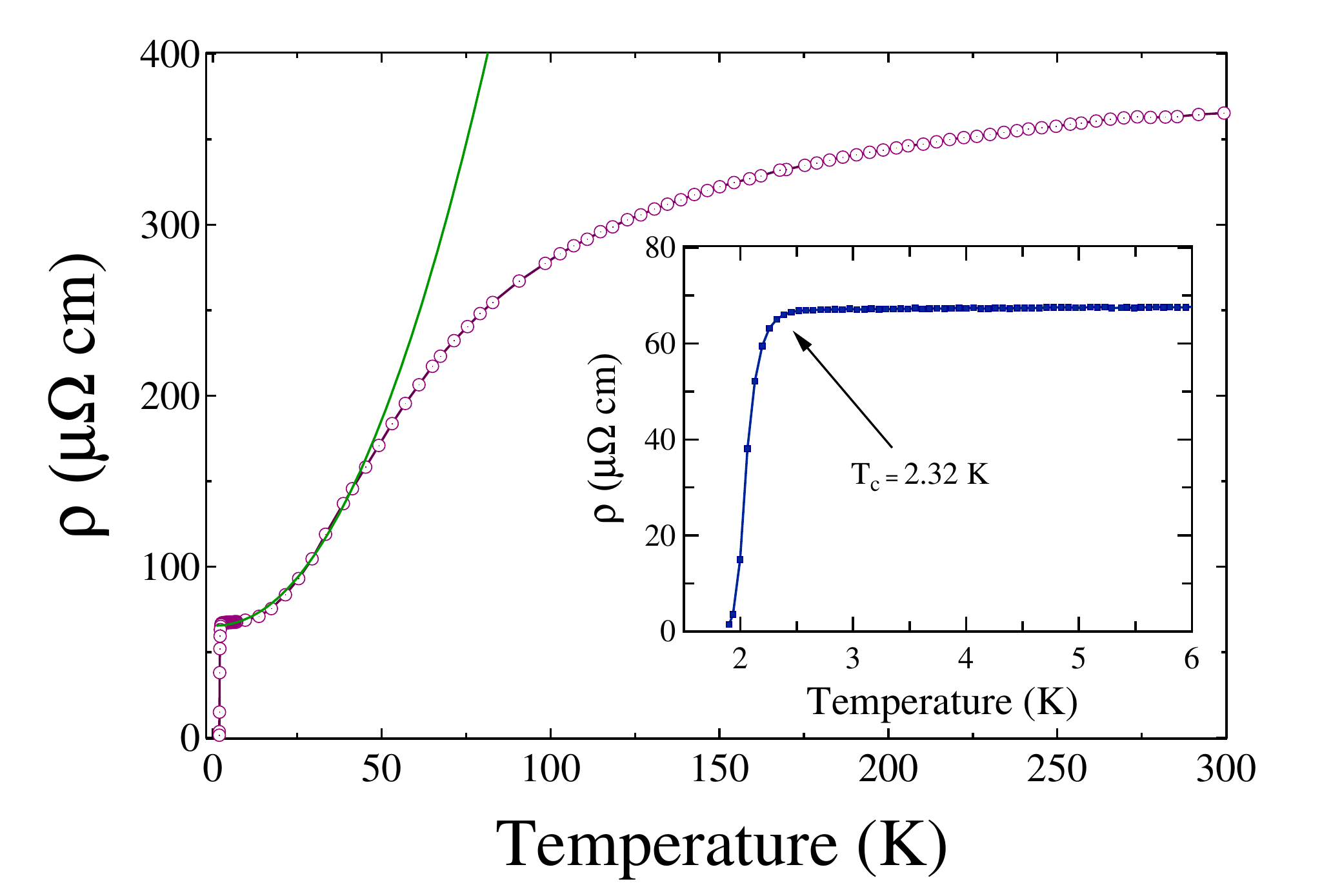}
\caption{Resistivity measurement taken against temperature $ \rho $(T) at zero field. The inset shows the drop in resistivity at T$ _{c} $ = 2.3 K. The green line is a fit to low temperature data using power law.}
\label{fig2}
\end{figure}

Figure \ref{fig2} shows the resistivity of Zr$ _{3} $Ir as a function of temperature. The transition temperature, observed from the resistivity measurement was around T$ _{c}^{onset} $ = 2.32 $ \pm $ 0.03 K. The low temperature (T$ _{c} $ $ < $ T $ \ll $ $ \theta_{D} $,) resistivity data show power law behaviour where $ \theta_{D} $ is the Debye temperature determined from specific heat measurement.  The inset shows an expanded plot of resistivity around the transition temperature. 
Low temperature resistivity data in the range 5 K $ \leq $ T $ \leq $ 45 K can be fitted fairly well with the equation
\begin{equation}
\rho(T) =  \rho_{0} + AT^{n}
\label{eq1}
\end{equation}

 \begin{figure*}[t]
\includegraphics[width=2.0\columnwidth,origin=b]{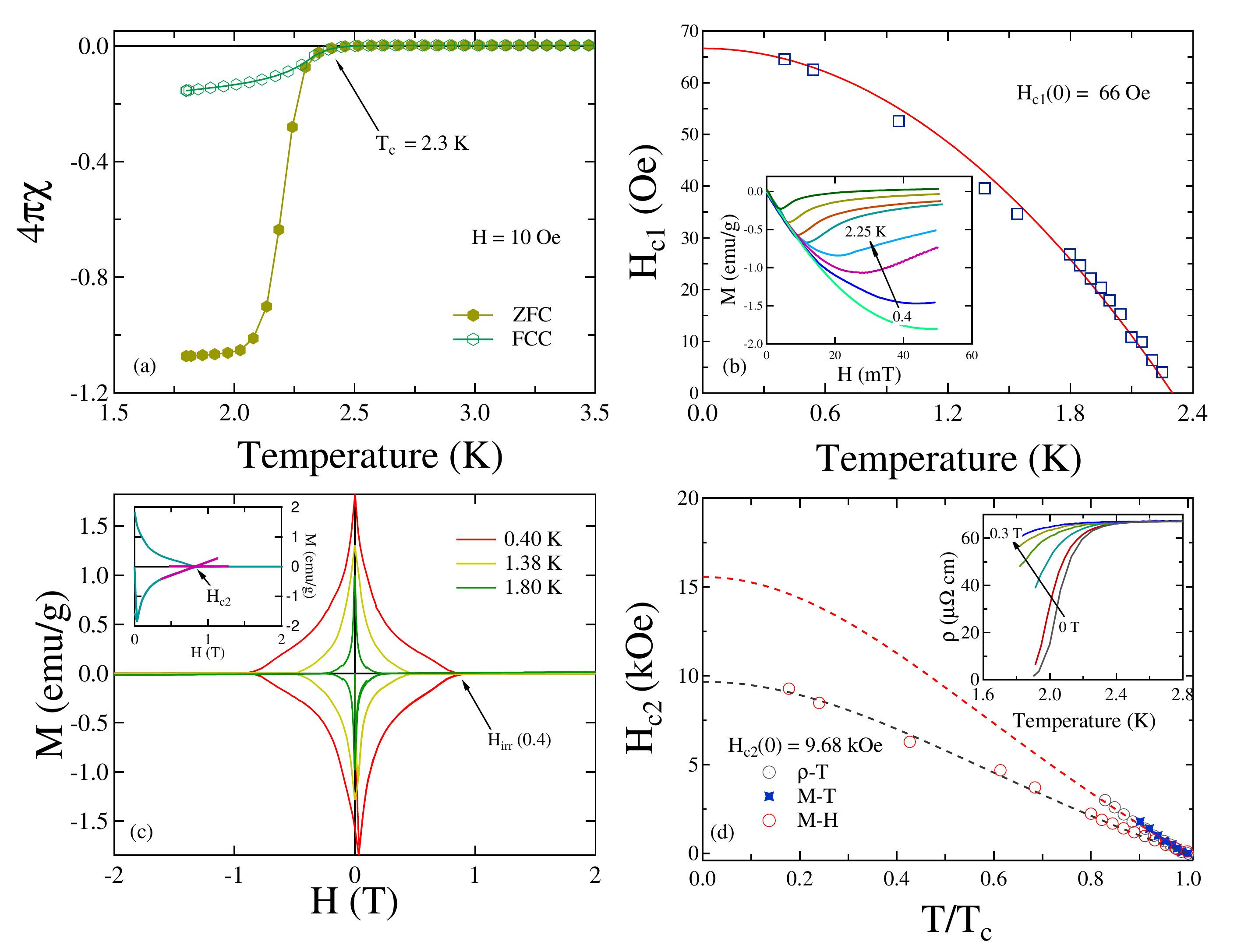}
\caption{(a)Temperature dependence of magnetic susceptibility collected via both ZFC and FCC mode. Superconducting transition of the sample is observed at T$ _{c} $ = 2.3 K. (b) Temperature variation of lower critical field H$ _{c1} $ collected via magnetization measurement. Extrapolation using G-L equation gives H$ _{c1} $(0) = 66 Oe. The inset shows the magnetization curves taken up to 500 Oe at different temperatures. (c) Isothermal magnetization curve taken at different temperatures. The inset shows the enlarged view. H$ _{c2} $ is taken at the discontinuity in the gradient as marked in the inset. (d) Determination of upper critical field H$ _{c2} $(0) using magnetization and resistivity measurements. The dotted lines indicates the fits to the data using Eq. \ref{HC2}. The inset shows low temperature resistivity data collected at different fields.}
\label{fig3}
\end{figure*}

Here $ \rho_{0} $ is the residual resistivity due to crystallographic defects and disorders while second term adds the electronic contribution to resistivity due to electron-electron correlation. Fitting with n = 2.1 yields the values as, $ \rho_{0} $ =  (64.9 $ \pm $ 0.1) $ \times $ $ 10^{-6} $ $ \Omega $ cm and A =  (0.047 $ \pm $ 0.006) $ \times $ $ 10^{-6} $ $ \Omega $ cm K$ ^{-2} $. A deviation from ideal Fermi behavior can be attributed to increased scattering in the system.

\subsection{Magnetization}
The DC magnetic susceptibility measured in both zero field cooled (ZFC) and  field cooled cooling (FCC) mode in an applied field of 10 Oe also confirms the bulk superconductivity with transition temperature T$^{onset} _{c} $ = 2.33 $ \pm $ 0.05 K (Fig. \ref{fig3}(a)). A superconducting fraction exceeding 100\% is accounted by the uncorrected geometrical factor. Fig. \ref{fig3}(b) depicts the low field magnetization data taken at different temperatures. The magnetization increases linearly with applied field after which it deviates due to vortex formation. The point of deviation from linear behaviour of the data is taken as the H$ _{c1} $ at each temperature. H$ _{c1} $(T) is modeled using the Ginzburg-Landau relation H$ _{c1} $(T) = H$ _{c1} $(0)(1-t$^{2}$), where t = T/T$ _{c} $, and we found H$ _{c1} $(0) = 66 $ \pm $ 4 Oe.

The upper critical field was estimated by magnetization as well as resistivity measurements at the different applied field in the range 100 Oe  $ \leq $ H $ \leq $ 3 kOe. The value of H$ _{c2} $ at each field is taken as the 90\% of the fall in moment/resistivity. It was seen from both the measurements that the T$ _{c} $ shifted towards lower value with a broader transition as field increases (Fig.  \ref{fig3}(d) inset). Magnetization data collected against the applied field at different temperature down to 0.4 K is shown in Fig. \ref{fig3}(c). It is visible from the figure that the area of hysteresis loop decreases as the temperature is increased towards T$ _{c} $, characteristic of type-II superconductor. A discontinuity in slope at a field, as shown in inset (Fig. \ref{fig3}(c)) is identified as  H$ _{c2} $ at each temperature. The value of H$ _{c2} $(0) is determined by fitting H$ _{c2} $(T) using the relation

\begin{equation}
H_{c2}(T) = H_{c2}(0)\frac{(1-t^{2})}{(1+t^2)} ,
\label{HC2}
\end{equation}

where t = T/T$ _{c} $ is the reduced temperature. Fitting the magnetization data using the equation yields H$ _{c2} $(0) = 9.68 $ \pm $ 0.42 kOe. The value of H$ _{c2} $(0) can be used to find Ginzburg-Landau coherence length using the equation
$ \xi_{GL} $ = ($\phi_{0}$/2$ \pi H _{c2}(0) $)$ ^{1/2} $, where $ \phi_{0} $ is the flux quantum ( $\phi_{0} $ = 2.07 $ \times $ 10$ ^{-15} $Tm$ ^{2} $). Substituting the value of H$ _{c2} $(0) gives $ \xi_{GL} $ = 185 $ \pm $ 4 \text{\AA}. Magnetic penetration depth for the sample $ \lambda_{GL}(0) $ is estimated using the relation 
\begin{equation}
H_{c1}(0) = \frac{\Phi_{0}}{4\pi\lambda_{GL}^2(0)}\left(\mathrm{ln}\frac{\lambda_{GL}(0)}{\xi_{GL}(0)}+0.12\right)   
\label{eqn6:ld}
\end{equation} 

which is obtained as 2624 $ \pm $ 68 \text{\AA}. This can be used to estimate Ginzburg-Landau parameter $ k $ = $ \lambda_{GL}(0)/\xi_{GL}(0) $ = 14.2  $ \pm $ 0.8, indicating the type II nature of the sample. \\
\\
One of the mechanism which causes the breaking of Cooper pair is the Pauli limiting field effect, in which the applied magnetic field induces the spin to align in the same direction of the magnetic field. The Pauli limiting field is estimated as H$ _{c2}^{p}(0) $ = 1.86T$ _{c} $ = 42.8 $ \pm $ 0.6 kOe.  This is larger than the H$ _{c2} $(0) obtained from magnetization as well as resistivity measurements. Another pair breaking mechanism is the orbital limiting field effect, $ H_{c2}^{orbital}(0) $, which can be  calculated by Werthermar-Helfand-Hohenberg expression \cite{WHH1,WHH2},

\begin{equation}
H_{c2}^{orbital}(0) = -\alpha T_{c}\left.\frac{dH_{c2}(T)}{dT}\right|_{T=T_{c}}
\label{eqn4:whh}
\end{equation} 

according to which the kinetic energy of the superelectron exceeds the condensation energy causing the Cooper pair breaking. Substituting $ \alpha $ = 0.693 for dirty limit superconductors (Dirty limit nature for Zr$ _{3} $Ir  is shown in the last section ) and initial slope at T = T$ _{c} $, $\frac{-dH_{c2}(T)}{dT} $ = 4.96 $ \pm $ 0.11 kOe/K gives $ H_{c2}^{orbital}(0) \approx $ 7.75 $ \pm $ 0.03 kOe. The values of H$ _{c2}^{p}(0) $ and $ H_{c2}^{orbital}(0) $ suggest that the orbital limiting field effect is the dominant pair breaking mechanism. A small value of the Maki parameter, $ \alpha_{M} $ = $ \sqrt{2}H_{c2}^{orb}(0)/H_{c2}^{p}(0)$ = 0.26 $ \pm $ 0.01 also indicates a negligible effect of Pauli limiting field.
\begin{figure} 
\includegraphics[width=9. cm, height=6. cm, origin=b]{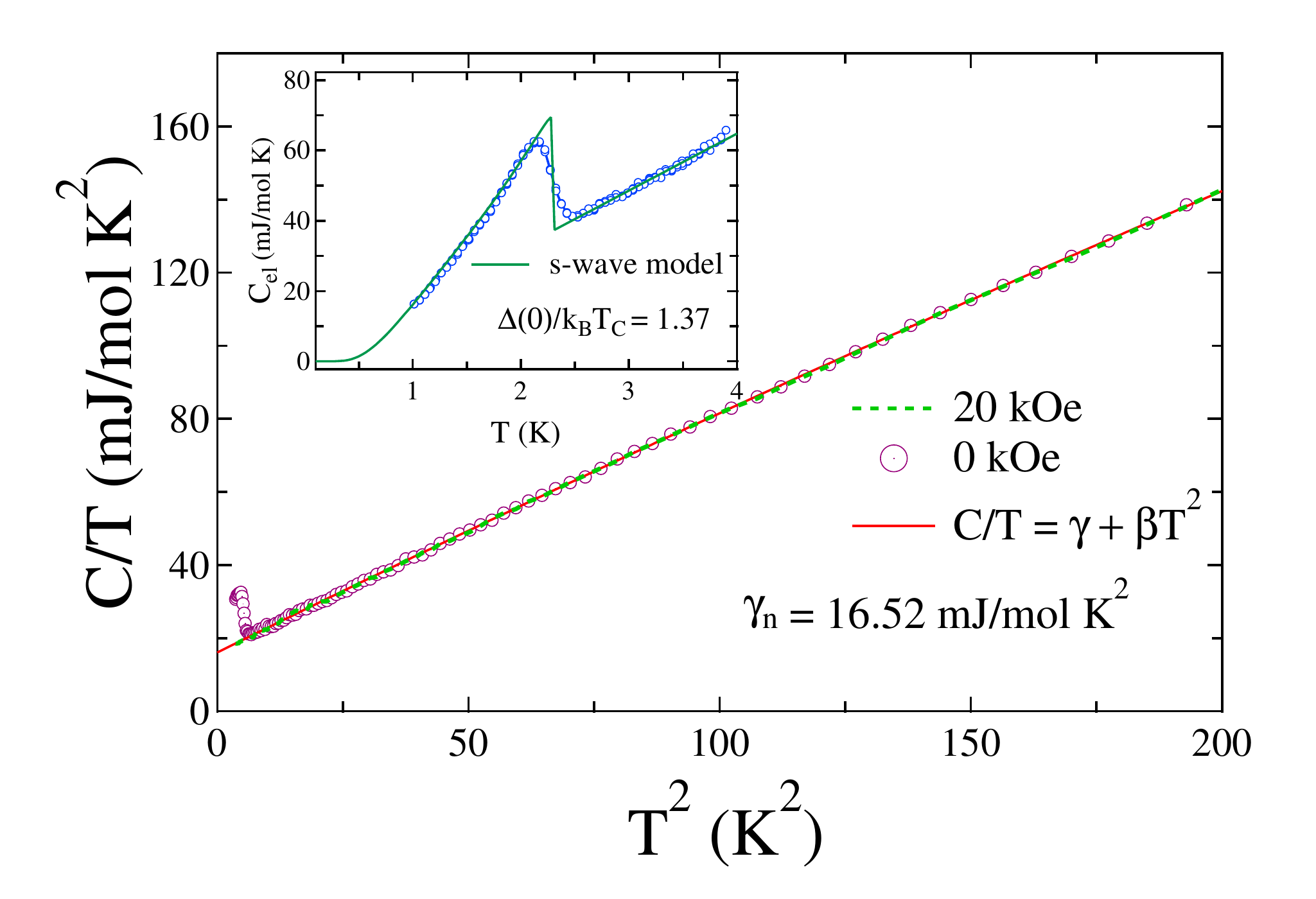}
\caption{C/T Vs T$ ^{2} $ shows a jump in specific heat at 2.31 K. An applied field of 20 kOe (green dotted line) which is above H$ _{C2} $(0) kills the superconducting  transition. (Inset) Variation of electronic specific heat with temperature is fitted by isotropic s-wave model (green line). The fitting yields the value of superconducting gap as $ \frac{\Delta(0)}{k_{B} T_{c}} $ = 1.37 }
\label{fig4}
\end{figure}

\subsection{Specific Heat}
Heat capacity measurements with temperature confirm the superconducting transition at T$ _{c} $ = 2.31 $ \pm $ 0.05 K. Transition temperature is consistent with the magnetization and resistivity data. Specific heat data was taken in the range 1 K $ \leq $ T $ \leq $ 15 K and normal state specific data was fitted using the relation C/T = $ \gamma_{n} $  +  $ \beta $T$^{2}  $ which allows the determination of the electronic contribution to specific heat $ \gamma_{n} $ = 16.52 $ \pm $ 0.09 mJ/mol K$ ^{2} $ and the phononic contribution, $ \beta $ = 0.618 $ \pm $ 0.001 mJ/mol K$ ^{4} $ (see Fig. \ref{fig4})

The electronic contribution to specific heat is calculated by subtracting the phononic contribution from the total specific heat (C$ _{el} $ = C $ - $ $ \beta $T$^{3}  $).  The normalised jump in specific heat, $ \frac{\Delta C_{el}}{\gamma_{n} T_{c}} $ came out to be 0.97 $ \pm $ 0.05 which is less than the BCS value of   $ \frac{\Delta C_{el}}{\gamma_{n} T_{c}} $ = 1.43 in the weak coupling limit. Similar low value of specific heat jump is reported for many other noncentrosymmetric materials \cite{LMP,FS}. This can be attributed to inhomogeneity in the sample or due to the presence of regions which does not participate in superconductivity.  \\
\\
Debye temperature of the sample $ \theta_{D} $ can be calculated using the value of $ \beta $ with the equation $ \theta_{D}= (12\pi^{4}RN/5\beta)^{\frac{1}{3}} $. Substituting the value of R, the universal gas constant, and number of atoms per formula unit cell N = 4 gives $ \theta_{D} $ = 232 $ \pm $ 8 K. Electronic density of states at the Fermi level $ D_{C}(E_{f})$ is proportional to the Sommerfeld coefficient $\gamma_{n}$. This can be calculated using the relation $\gamma_{n}= (\pi^{2}k_{B}^{2}/3)D_{C}(E_{f})$, which gives $ D_{C}(E_{f}) $ = 7.01 $ \pm $ 0.03 $ \frac{states}{eV f.u} $ . Once the Debye temperature $ \theta_{D} $ is determined, one can use McMillans Eq. \ref{eqn8} \cite{Mc} to  evaluate the electron phonon coupling constant, a dimensionless number which says about the relative strength of electron-phonon coupling 
\begin{equation}
\lambda_{e-ph} = \frac{1.04+\mu^{*}ln(\theta_{D}/1.45T_{c})}{(1-0.62\mu^{*})ln(\theta_{D}/1.45T_{c})-1.04 }
\label{eqn8}
\end{equation} 
This provides us $ \lambda_{e-ph} $ = 0.56 $ \pm $ 0.04 indicating Zr$ _{3} $Ir is an intermediately coupled superconductor. The bare-band structure density of states, $ D_{band}(E_{f}) $ is related to electron-phonon coupling strength $ \lambda_{e-ph} $ and can be calculated using
 $D_{C}(E_{f}) = D_{band}(E_{f})(1+\lambda_{e-ph})$ which gives $ D_{band}(E_{f}) $ = 4.5 $ \pm $ 0.3  $ \frac{states}{eV\textit{ f}.u} $. The effective mass m$ ^{*} $ of the quasi-particle came out to be 1.56m$ _{e} $ using the relation $ m^{*} = m^{*}_{band}(1+\lambda_{e-ph}) $, where we have used m$ ^{*}_{band} $ = m$ _{e} $.

The behaviour of electronic specific heat gives salient features of superconducting gap structure. The temperature dependence of electronic specific heat $ C_{el} $ is shown in Fig. \ref{fig4}. The superconducting contribution to entropy (S) as described by BCS theory is 
\begin{equation}
\frac{S}{\gamma_{n}T_{c}} = -\frac{6}{\pi^2}\left(\frac{\Delta(0)}{k_{B}T_{c}}\right)\int_{0}^{\infty}[ \textit{f}\ln(f)+(1-f)\ln(1-f)]dy \\
\label{eqn10:s}
\end{equation} 

where the integral is taken over the energies of normal electrons relative to the Fermi level.  $\textit{f}$($\xi$) = [exp($\textit{E}$($\xi$)/$k_{B}T$)+1]$^{-1}$ is the Fermi-Dirac distribution function. The energy of the quasiparticle is given by $\textit{E}$($\xi$) = $\sqrt{\xi^{2}+\Delta^{2}(t)}$, where $\textit{y}$ = $\xi/\Delta(0)$, $\mathit{t = T/T_{c}}$ and $\Delta(t)$ = tanh[1.82(1.018(($\mathit{1/t}$)-1))$^{0.51}$] is the BCS approximation for the temperature dependence of energy gap. The normalised electronic specific heat can be related to normalised entropy by 
\begin{equation}
\frac{C_{el}}{\gamma_{n}T_{c}} = t\frac{d(S/\gamma_{n}T_{c})}{dt} \\
\label{eqn11:Cel}
\end{equation}
where $ C_{el} $ below T$ _{c} $ is described by the above equation whereas above $ T_{c} $ its equal to $ \gamma_{n}$T$_{c} $. The data was fitted quite well with the equation \ref{eqn11:Cel}. This yielded the value of the superconducting energy gap as  $ \frac{\Delta(0)}{k_{B} T_{c}} $ = 1.37 $ \pm $ 0.04. The value obtained is below the BCS predicted value $ \frac{\Delta(0)}{k_{B} T_{c}} $ = 1.76. 

\begin{figure} 
\includegraphics[width=9. cm, height=6. cm, origin=b]{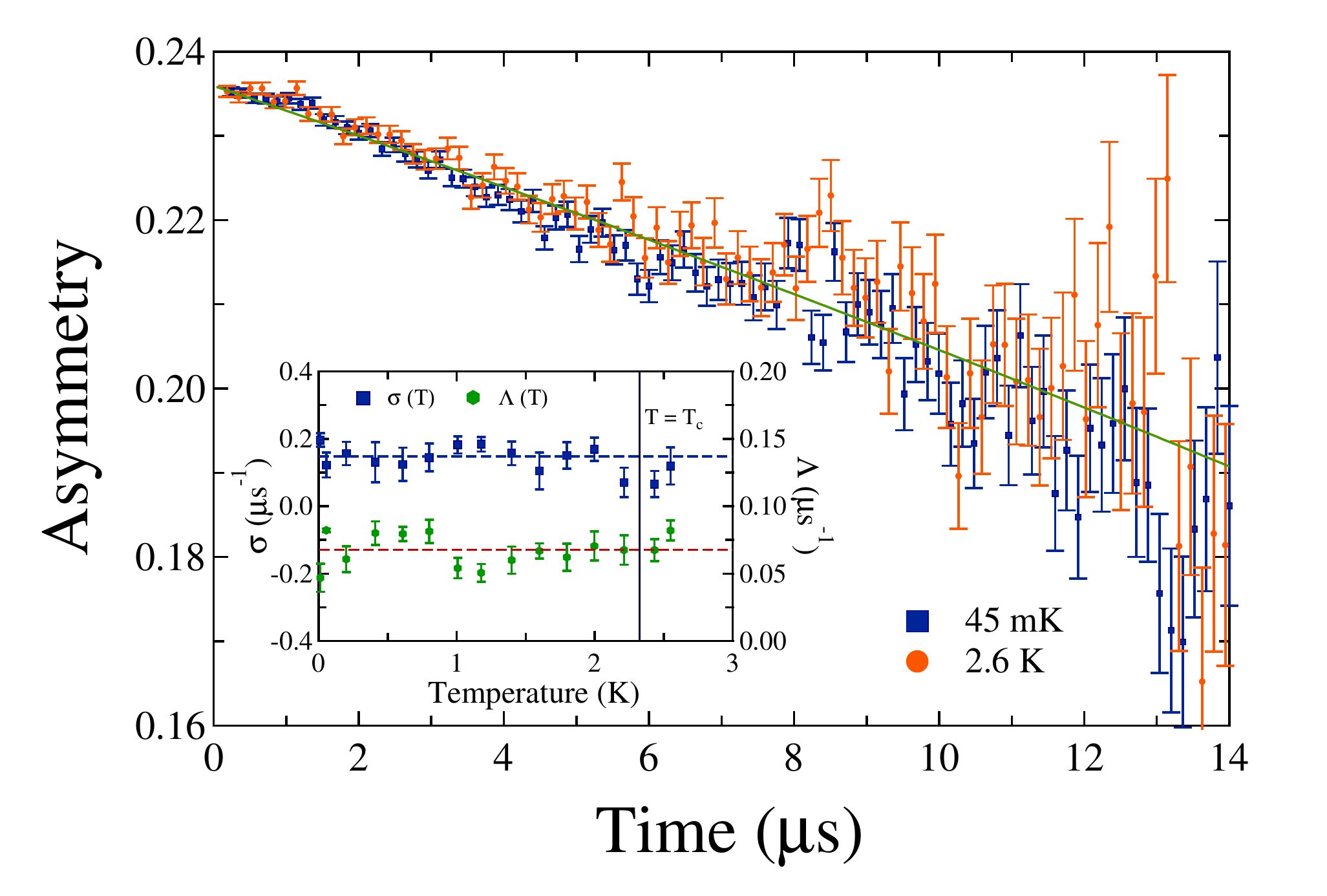}
\caption{$ \mu $SR spectra collected in zero field configuration at temperatures above (2.6 K) and below (45 mK) the transition temperature. The inset shows no significant change in fit parameters $ \Lambda $(T) and $ \sigma $(T) across the transition temperature. }
\label{fig5}
\end{figure}

\subsection{Muon Spin Relaxation and Rotation}
A further investigation of the superconducting ground state was undertaken by muon spin relaxation and rotation measurements. Zero field muon spin relaxation spectra were collected at different temperatures above and below the T$ _{c} $. Figure \ref{fig5} shows the representative spectra at 2.6 K and 45 mK. The absence of any oscillatory component in the data rules out the presence of any spontaneous coherent field associated with the ordered magnetic structure. In the absence of any coherent magnetic ordering, muon spin relaxation is determined primarily by randomly oriented local nuclear dipole moments, which can be modeled by the Gaussian Kubo-Toyabe (KT) function 
\begin{equation}
G_{\mathrm{KT}}(t) = \frac{1}{3}+\frac{2}{3}(1-\sigma^{2}_{\mathrm{ZF}}t^{2})\mathrm{exp}\left(\frac{-\sigma^{2}_{\mathrm{ZF}}t^{2}}{2}\right) 
\label{eqn17:zf}
\end{equation} 

where $\sigma _{ZF} $ corresponds to the relaxation due to static, randomly oriented local fields associated with the nuclear moments at the muon site. The zero field spectra of Zr$ _{3} $Ir can be well described by the function
\begin{equation}
A(t) = A_{1}G_{\mathrm{KT}}(t)\mathrm{exp}(-\Lambda t)+A_{\mathrm{BG}} 
\label{eqn18:tay}
\end{equation}
where A$ _{1} $ corresponds to the sample asymmetry, $ \Lambda $ is the additional electronic relaxation rate, and A$ _{BG} $ is the temperature independent background asymmetry coming from the muons stopped in the silver sample holder. Reportedly, in superconducting systems where time-reversal symmetry is broken, spontaneous magnetic moments arise below T$ _{c}$, and an increase may be observed in either $\sigma _{ZF} $ or $ \Lambda $. Fitting Eq. \ref{eqn18:tay} at different data sets above and below T$ _{c}$ yields similar values of $\sigma _{ZF} $ and $ \Lambda $ with no noticeable change (see inset Fig. \ref{fig5}). This suggests that the time reversal symmetry is preserved in the superconducting phase.

\begin{figure} 
\includegraphics[width=9. cm, height=6. cm, origin=b]{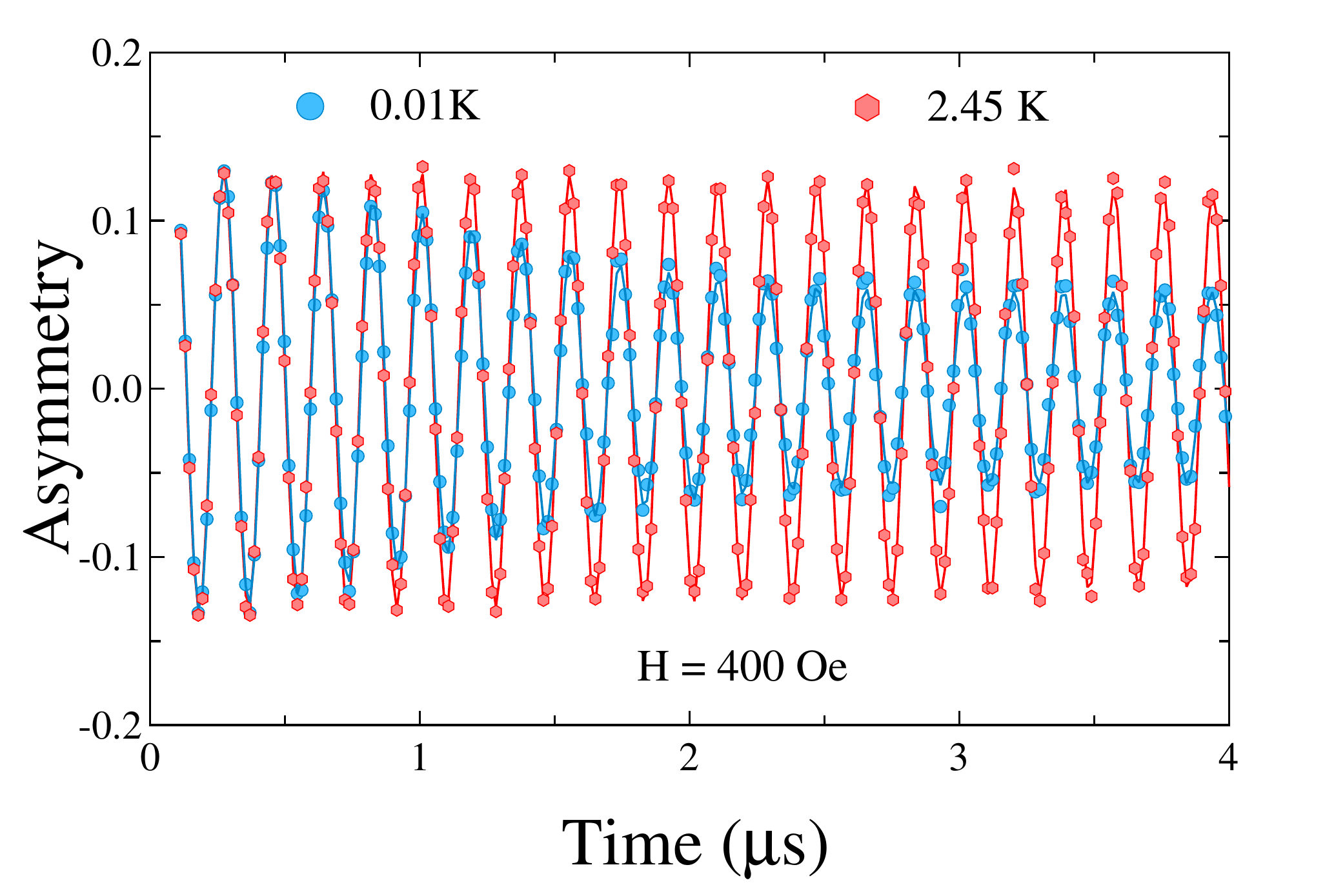}
\caption{A representative $ \mu $SR spectra collected in transverse field configuration at an applied field of 400 Oe. The spectra collected at 0.01 K show significant decay due to flux line lattice formation, while the non decaying nature of signal at T = 2.45 K indicates uniform field distribution above T$ _{c} $ }
\label{fig6}
\end{figure}

Superconducting gap structure of Zr$ _{3} $Ir was examined using transverse field $ \mu $SR (TF-$ \mu $SR). TF-$ \mu $SR precession signal was collected at 400 Oe, which is well above the lower critical field. During the experiment, the sample was field cooled to the low temperature to ensure the formation of a well-ordered flux line lattice (FLL). Figure \ref{fig6} shows the muon spin rotation spectra measured at either side of transition temperature T$_{c}$. It is quite evident from the graph that the quick decay in spectra below T$ _{c} $  is accounted by the formation of FLL, which causes inhomogeneous field distribution. TF-$ \mu $SR signal is well described by the oscillatory decaying function given by
\begin{equation}
G_{\mathrm{TF}}(t) = A_{1}\mathrm{exp}\left(\frac{-\sigma^{2}t^{2}}{2}\right)\mathrm{cos}(w_{1}t+\phi)+A_{2}\mathrm{cos}(w_{2}t+\phi) ,
\label{eqn19:Tranf}
\end{equation}

where $ \omega_{1} $ = $ \gamma_{\mu}B_{1} $ and $ \omega_{2} $ = $ \gamma_{\mu}B_{2} $ are the muon precessional frequencies for sample and background respectively, $ \phi $ is the initial phase offset, $ \sigma $ is the total depolarization rate, and $ \gamma_{\mu}$ is muon gyromagnetic ratio. The field distribution due to FLL is broadened by the presence of randomly oriented nuclear magnetic moments. Hence the total depolarization rate $ \sigma $ is written as

\begin{equation}
\sigma^{2} = \sigma_{\mathrm{N}}^{2}+\sigma_{\mathrm{FLL}}^{2}
\label{eqn19:sigma}
\end{equation}

Here $ \sigma_{\mathrm{N}} $ corresponds to depolarization due to nuclear moments and $ \sigma_{\mathrm{FLL}} $ corresponds to that from FLL.

\begin{figure} 
\includegraphics[width=9. cm, height=6. cm, origin=b]{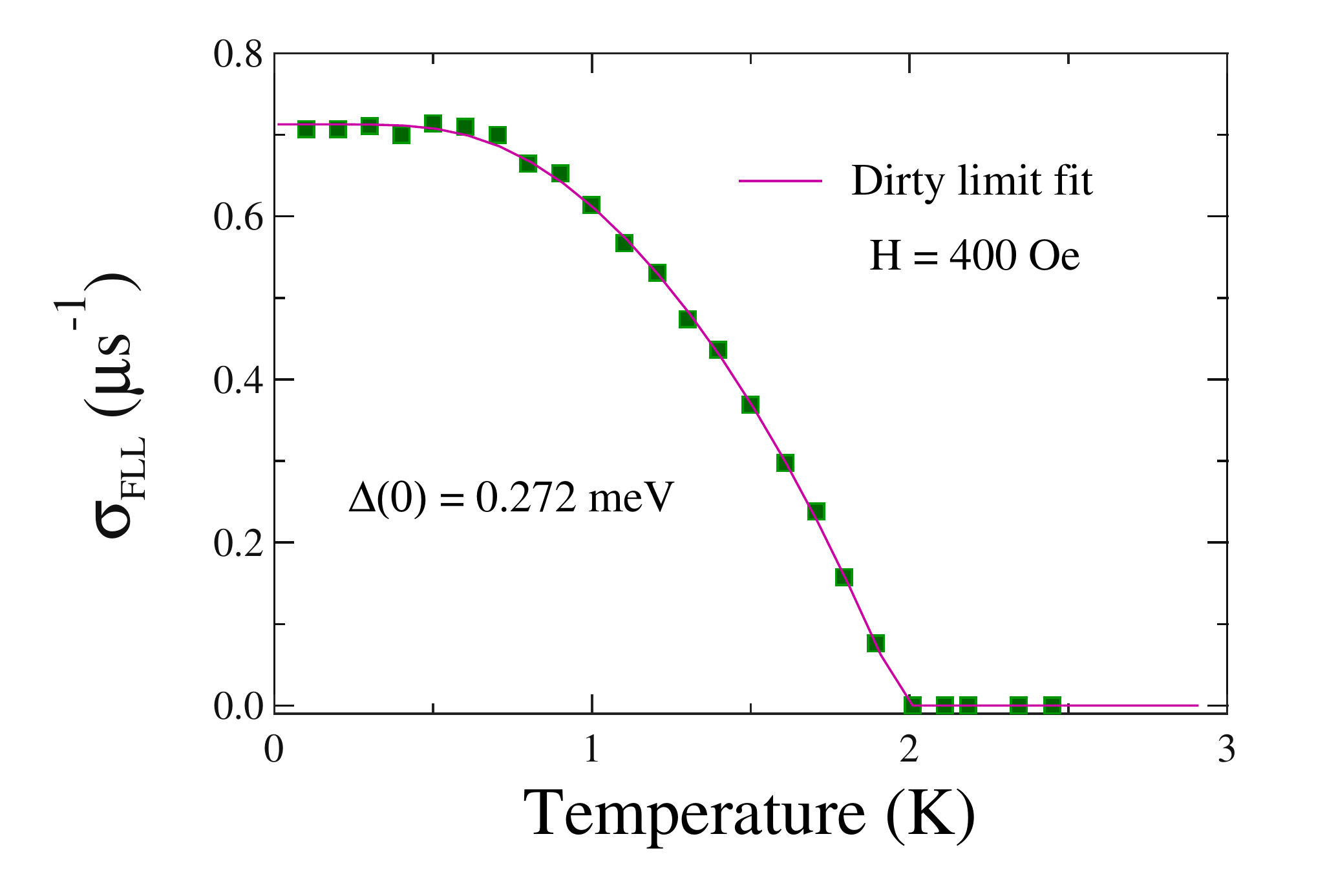}
\caption{Temperature dependence of $ \sigma_{FLL} $ measured at an applied field of 400 Oe. Solid red line is the dirty limit isotropic s-wave fit for the data.}
\label{fig7}
\end{figure}

According to the model which explains dirty limit (see next page) s-wave superconductors, the temperature dependence of $ \mu $SR depolarization rate in the vortex state can be written as
\begin{equation}
\frac{\sigma_{FLL}(T)}{\sigma_{FLL}(0)} = \frac{\Delta(T)}{\Delta(0)}\mathrm{tanh}\left[\frac{\Delta(T)}{2k_{B}T}\right] ,
\label{eqn22:lpd}
\end{equation}

where $\Delta$(T) = $\Delta_{0}$$\delta(T/T_{c})$ and $\delta(T/T_{c})$ = tanh[1.82(1.018($\mathit{(T_{c}/T})$-1))$^{0.51}$] is the BCS approximation for temperature dependence of superconducting energy gap.
Combining Eq. \ref{eqn19:sigma} and Eq. \ref{eqn22:lpd}, a model is obtained where $ \sigma $(T) below T$ _{c} $ is well described by the relation 
\begin{equation}
\sigma(T) = \sqrt{\sigma_{\mathrm{FLL}}^{2}(0)\frac{\Delta^{2}(T)}{\Delta^{2}(0)}\mathrm{tanh}^{2}\left[\frac{\Delta(T)}{2k_{B}T}\right]+\sigma_{\mathrm{N}}^{2}}
\label{eqn23:fs}
\end{equation}
whereas above T$ _{c} $ it is simply equal to $ \sigma_{N} $. Figure \ref{fig7} represents temperature dependent depolarization rate due to FLL  at 400 Oe, calculated using Eq. \ref{eqn19:sigma}. The depolarization rate $ \sigma_{FLL} $(T) is zero as expected above T$ _{c} $. The best fit to the data using Eq. \ref{eqn22:lpd} gives  $ \Delta $(0) = 0.272 meV. This gives the normalized energy gap at T$ _{c} $,  $ \frac{\Delta(0)}{k_{B} T_{c}} $ = 1.372, which accurately matches with the results from specific heat measurement.\\

\begin{figure}
\includegraphics[width=9. cm, height=6. cm, origin=b]{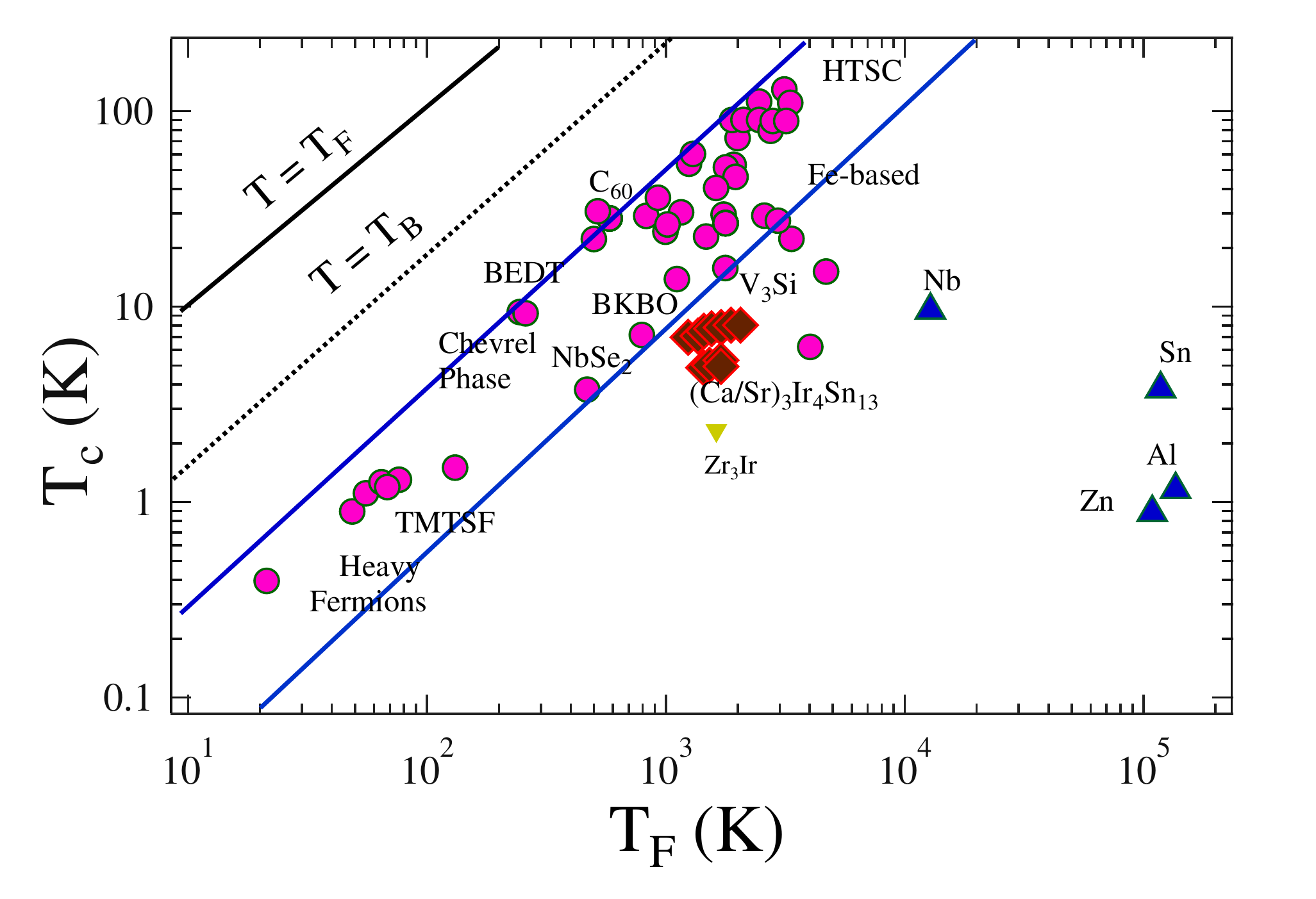}
\caption{\label{fig8} The Uemura plot showing the superconducting transition temperature $T_{c}$ vs the effective Fermi temperature $T_{F}$, where Zr$ _{3} $Ir is shown as an inverted yellow triangle. Other data points plotted between the blue solid lines is the different families of unconventional superconductors \cite{KKC,RKH}.} 
\end{figure}

The results from the bulk measurements can be effectively used to determine more parameters for Zr$ _{3} $Ir which characterizes superconducting ground state. A set of four equations as explained in Ref. \cite{RZ2,AD} is solved to get superconducting carrier density n, effective mass m$ ^{*} $, BCS coherence length $ \xi_{0} $, mean free path $ l $ and are tabulated in Table \ref{elec propr}.
 Solving these has confirmed that sample is in the dirty limit regime ( $ \xi_{0} $ > \textit{l$_{e}$} ) justifying the dirty limit model used in both specific heat and muon spectra analysis. The Fermi temperature (T$ _{F} $) of the sample can be extracted from the relation
\begin{equation}
 k_{B}T_{F} = \frac{\hbar^{2}}{2}(3\pi^{2})^{2/3}\frac{n^{2/3}}{m^{*}}, 
\label{eqn13:tf}
\end{equation} 

which gives the effective Fermi temperature T$ _{F} $ = 1645 K. In general, high T$ _{c} $ superconductors and other unconventional superconductors falls in the range 0.01 $ \leq $ T$ _{c} $/T$ _{F} $ $ \leq $ 0.1 \cite{F1,F2,F3}. For Zr$ _{3} $Ir the ratio $ \frac{T_{c}}{T_{F}} $ comes around 0.0014 which places our sample away from the unconventional family as shown in Fig. \ref{fig8}.

At this point, it will be worth to discuss the dirty limit nature of the sample which can have implications on the measurements determining the superconducting gap structure, as done in \cite{CIS2}. A superconductor in the dirty limit regime will have an increased scattering from impurities and defects. This effects can have impacts on different measurements at low temperatures, which in turn can give slightly different results, suppressing nodal or anisotropic behavior. A notable example is the case of Mg$ _{10} $Ir$ _{19} $B$ _{16} $ where a sample with residual resistivity 1400 $ \mu\Omega $cm \cite{Mg1} has shown no evidence of unconventional gap structure. While a comparatively clean sample of Mg$ _{10} $Ir$ _{19} $B$ _{16} $ with a residual resistivity value of 100 $ \mu\Omega $cm \cite{Mg2} has shown a two gap behaviour at low temperature penetration depth study using tunnel diode oscillator. Though most of the superconductors which have shown non-isotropic behaviour like CePt$ _{3} $Si, Li$ _{2} $Pt$ _{3} $B falls in the clean limit regime, many noncentrosymmetric superconductors with considerable residual resistivity value has shown an isotropic s-wave behaviour. Considering the present case of Zr$ _{3} $Ir, the sample has shown a residual resistivity value of 61 $ \mu\Omega $cm, which is comparatively less and the depolarization rate is seen temperature independent below 0.7 K within the statistical uncertainty. This strongly suggests that the sample has an isotropic gap, and any other possibility can be ruled out.\\

\begin{table}[h!]
\caption{Normal and superconducting properties of noncentrosymmetric superconductor Zr$_{3}$Ir}
\label{elec propr}
\begin{center}
\begin{tabular*}{1.0\columnwidth}{l@{\extracolsep{\fill}}lll}\hline\hline
Parameter& unit& value\\
\hline
\\[0.5ex]                                        
$T_{c}$& K& 2.33\\             
$H_{c1}(0)$& Oe& 66\\                       
$H_{c2}(0)$& kOe& 9.68\\
$H_{c2}^{P}(0)$& kOe& 42.8\\
$\xi_{GL}$& \text{\AA}& 185\\
$\lambda_{GL}$& \text{\AA}& 2624\\
$\kappa_{GL}$& &14.2\\
$\Delta C_{el}/\gamma_{n}T_{c}$&   &0.97\\
$\Delta(0)/k_{B}T_{c}$&  &1.37\\
$m^{*}/m_{e}$& & 14.8\\             
n& 10$^{27}$m$^{-3}$& 13.8\\
$l$&  \text{\AA}& 34.66\\ 
$\xi_{0}$&  \text{\AA}& 46\\                      
$\xi_{0}/l$& & 1.33\\
$v_{f}$& 10$^{4}$ms$^{-1}$& 5.81\\
$\lambda_{L}$& \text{\AA}& 1739\\
$T_{c}$/$T_{F}$& &0.0014\\
\\[0.5ex]
\hline\hline
\end{tabular*}
\par\medskip\footnotesize
\end{center}
\end{table}

\section{CONCLUSIONS}
In conclusion, we have studied the superconducting properties of Zr$ _{3} $Ir, which belong to the tetragonal noncentrosymmetric $ \alpha$-$V_{3}S$ family. This is one of the first superconductor of this family. The transport, magnetization and specific heat measurements confirms a type-II superconductivity with  T$ _{c} $ = 2.3 K. The lower and upper critical field value is estimated as H$ _{c1} $ = 66 $ \pm $ 4 Oe and H$ _{c2} $ = 9.68 $ \pm $0.42 kOe. The characteristic length scale for the compound is estimated using the G-L relations which came out to be $ \xi_{GL} $ = 185 $ \pm $ 4 \text{\AA} and $ \lambda_{GL} $ = 2624 $ \pm $ 68 \text{\AA}. The value of the normalised specific heat jump  $ \frac{\Delta C_{el}}{\gamma T_{c}} $ = 0.97 $ \pm $ 0.05 along with the superconducting region fitted using the weak coupling limit BCS expression suggested that Zr$ _{3} $Ir is a s-wave superconductor with isotropic gap $ \frac{\Delta(0)}{k_{B} T_{c}} $ = 1.37 . This was further confirmed by TF-$ \mu $SR measurements. $ \mu $SR measurements also rule out the presence of any spontaneous magnetic field arising in the superconducting state, confirming the preserved time reversal symmetry for the system in the sensitivity limit of the muon. To understand the role of the structure/ pairing symmetry in noncentrosymmetric superconductors, further experimental work coupled with theoretical calculations is vital. This work paves the way for further studies on new members from $\alpha$-V$_{3}$S family of compounds to understand the role of the different crystal structure (noncentrosymmetric)/ spin orbital coupling on time reversal symmetry breaking in noncentrosymmetric superconductors. 

\section{ACKNOWLEDGMENTS}
R.P.S. acknowledges Science and Engineering Research Board, Government of India for the Young Scientist Grant No. YSS/2015/001799 and Financial support from DST FIST Project No. SR/FST/PSI-195/2014(C) is also thankfully acknowledged. We thank ISIS, STFC, UK for the beamtime to conduct the $\mu$SR experiments \cite{ZIM}.

\end{document}